\documentclass[preprint,dvips,dvipdfm,sort&compress]{elsarticle}

\usepackage{graphicx}
\usepackage{graphics}
\usepackage{color}
\usepackage{amsmath}
\usepackage{amssymb}
\usepackage{natbib}

\begin{document}

\begin{frontmatter}

\title{Enhanced collectivity in $^{74}$Ni}

\author[riken]{N.~Aoi}\ead{aoi@riken.jp}
\author[rikkyo,riken]{S.~Kanno}
\author[riken]{S.~Takeuchi}
\author[tokyo]{H.~Suzuki}
\author[msu]{D.~Bazin}
\author[msu]{M.~D.~Bowen}
\author[msu]{C.~M.~Campbell}
\author[msu]{J.~M.~Cook}
\author[msu]{D.-C.~Dinca}
\author[msu]{A.~Gade}
\author[msu]{T.~Glasmacher}
\author[tokyo,msu]{H.~Iwasaki}
\author[riken]{T.~Kubo}
\author[rikkyo]{K.~Kurita}
\author[riken]{T.~Motobayashi}
\author[msu]{W.~F.~Mueller}
\author[toko]{T.~Nakamura}
\author[tokyo,riken]{H.~Sakurai}
\author[medi,rcnp]{M.~Takashina}
\author[msu]{J.~R.~Terry}
\author[msu,riken]{K.~Yoneda}
\author[msu]{H.~Zwahlen}

\address[riken]
{RIKEN Nishina Center,  
2-1 Hirosawa, Wako, Saitama 351-0198, Japan}
\address[rikkyo]
{Department of Physics, Rikkyo University,
3-34-1 Nishi-Ikebukuro, Toshima, Tokyo 171-8501, Japan}
\address[msu]
{National Superconducting Cyclotron Laboratory, Michigan State University,
East Lansing, Michigan 48824-1321, USA}
\address[tokyo]
{Department of Physics, University of Tokyo,
7-3-1 Hongo, Bunkyo, Tokyo 113-0033, Japan}
\address[toko]
{Department of Physics, Tokyo Institute of Technology,
2-12-1 Oh-okayama, Meguro-ku, Tokyo 152-8551, Japan}
\address[medi]
{Graduate School of Medicine, Osaka University,
Suita, Osaka 565-0871, Japan}
\address[rcnp]
{Research Center for Nuclear Physics,
Osaka University, Ibaraki, Osaka 567-0047, Japan}

\begin{abstract}
The neutron-rich nucleus $^{74}$Ni was studied 
with inverse-kinematics inelastic proton scattering using a $^{74}$Ni
radioactive beam incident on a liquid hydrogen target 
at a center-of-mass energy of 80 MeV. 
From the measured de-excitation $\gamma$ rays, the population of the first $2^+$
state was quantified. The angle-integrated excitation cross section was
determined to be 14(4)~mb.  
A deformation length of $\delta =$ 1.04(16)~fm was extracted 
in comparison with distorted wave theory, which suggests that 
the enhancement of collectivity established for $^{70}$Ni continues up to
$^{74}$Ni. A comparison with results of shell model and quasi-particle 
random phase approximation calculations indicates that 
the magic character of $Z=28$ or $N=50$ is weakened in $^{74}$Ni. 
\end{abstract}


\end{frontmatter}

\section{Introduction}
\label{sec1}
The evolution of the nuclear shell structure is one of the most fundamental 
topics in the investigation of unstable nuclei. Recent studies have shown that 
the shell structure established for stable nuclei can change 
in nuclei far from the valley of $\beta$-stability~\cite{magic_review}.  
Some of the traditional magic numbers lose their magicity in 
neutron-rich regions, while some other proton or neutron numbers 
emerge as magic. For example, the neutron-rich nuclei $^{12}$Be and $^{32}$Mg 
were found to have large collectivity 
in spite of their neutron numbers $N=8$ and $N=20$, respectively, which are
well-established magic numbers in stable
nuclei~\cite{Navin,Iwasaki,12Be_iwasaki,12Be_imai,Detraz,32Mg_Ex,32Mg_motobaya}.
It is interesting to extend the study to heavier regions of the nuclear chart  
to investigate whether such structural changes occur there as well. 
One particular interest is the behavior of the 
magicity at proton number $Z=28$ and neutron number $N=50$ 
in the neutron-rich nucleus $^{78}$Ni and neighbouring isotopes. 
In order to address this question, we have studied the collectivity of 
$^{74}$Ni with $Z=28$ and $N=46$ by measuring the proton inelastic scattering 
to the first excited $2^+$ ($2^+_1$) state. 

\begin{figure}[t]
 \includegraphics[keepaspectratio, scale=0.7]{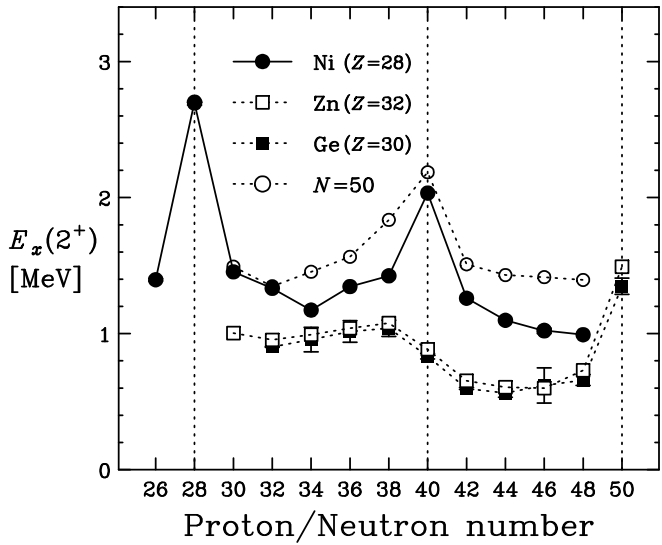}
 \caption{Experimental $E_{\rm x}(2^{+})$ values for Ni isotopes ($Z=28$) 
   and the neighbouring even-$Z$ isotopes, Zn ($Z=30$) and  
   Ge ($Z=32$), as a function of the neutron number. 
   The $E_{\rm x}(2^{+})$ values for the $N=50$ isotones are also shown 
   as a function of proton number. The values are taken from 
   Refs.~\cite{Yur04-1,Yam05,Ram01,72Ni_sawicka,Maz05,Van07}. 
 \label{fig:e2-n50niznge}}
\end{figure}

The Ni isotopes have reduced collectivity reflecting the magic character of 
the closed proton shell at $Z=28$. The excitation energies of the $2^+_1$ states
($E_{\rm x}(2^{+}_{1})$) along the Ni isotopic chain are 
higher than those of the neighbouring even-$Z$ isotopes such as Zn ($Z=30$) or 
Ge ($Z=32$) (Fig.~\ref{fig:e2-n50niznge}). 
However, for the Ni isotopes between neutron numbers $N=40$ and $N=50$, where
the valence neutrons occupy the $g_{9/2}$ orbital, the $E_{\rm x}(2^{+})$ value
decreases continuously. 
This feature does not match with the expectations that arise assuming $Z=28$ as
a good magic number. Single-closed-shell nuclei 
with valence nucleons in a single $j$ shell can be described 
in terms of a seniority scheme ~\cite{Casten, Talmi}, 
leading to constant $E_{\rm x}(2^{+})$ values along the isotopic chain. 
Indeed, for the $N=50$ isotones with valence protons in the $g_{9/2}$ orbital,
the $E_{\rm x}(2^{+})$ values are almost constant between 
$^{92}$Mo ($N=50, Z=42$) and $^{98}$Cd ($N=50, Z=48$) 
as shown by open circles in Fig.~\ref{fig:e2-n50niznge}. 
The monotonic decrease of $E_{\rm x}(2^{+})$ in the Ni isotopes 
up to $^{76}$Ni may be an indication of the magic character of 
$Z=28$ or $N=50$ being weakened.  

The degree of collectivity can be characterized more directly by 
the quadrupole deformation parameter $\beta_2$ or the deformation length $\delta$
which is the product of $\beta_2$ and the nuclear radius $R$. 
These quantities can be derived from the transition strength between the 
ground $0^+$ state and $2^+_1$ state using various probes, such as 
electromagnetic transition or proton inelastic scattering.
When extracting $\beta_2$ and $\delta$ from the cross sections of the 
proton inelastic scattering, the deformation length $\delta$ is less model
dependent~\cite{12Be_iwasaki} and subsequently used as a measure of
collectivity in this work.   
The $\delta$ value derived from the reduced electric quadrupole 
transition probability $B({\rm E2})$, which is known up to $^{70}$Ni, 
is very small in $^{68}$Ni~\cite{Sor02} and increases abruptly 
at $^{70}$Ni~\cite{Per06}. 
In Ref.~\cite{Per06} this is interpreted as an indication that
the shell gap at $Z=28$ in $^{70}$Ni is reduced 
by the occupation of the neutron $g_{9/2}$ orbital. 
Beyond $^{70}$Ni, where further enhancement of the collectivity is suggested 
by the decreasing $E_{\rm x}(2^{+}_1)$ value, $\delta$ is not known 
due to the difficulty of producing such neutron-rich nuclei 
at sufficient intensity for spectroscopy. 
In this work, we have succeeded to obtain the deformation length $\delta$ for
$^{74}$Ni from the angle-integrated cross section for the excitation of the
first $2^+$ state in the proton inelastic scattering.
 
The difficulty caused by the low $^{74}$Ni beam intensity has been overcome 
by incorporating an efficient combination of proton inelastic scattering 
and $\gamma$-ray spectroscopy. The excited states of $^{74}$Ni were identified by 
measuring the de-excitation $\gamma$ rays~\cite{12Be_iwasaki} 
instead of the recoil protons~\cite{Kra94,Alex95}. Consequently, 
a very thick proton target could be used without loss of energy resolution. 
Furthermore, by using a liquid hydrogen target, 
very highly sensitive measurements can be realized as demonstrated 
with the case of $^{30}$Ne~\cite{30Ne_Yanagisa}, because hydrogen has the 
largest number density of atoms within the target material with a given
energy-loss-equivalent thickness.  
High efficiency is also obtained thanks to the small scattering angle 
in the laboratory reference frame due to the use of inverse kinematics 
with the lightest-mass target, where 100\% acceptance for  
the scattered projectiles can easily be achieved in an experimental setup. 
Another advantage of using a liquid hydrogen target is the significant 
reduction of $\gamma$-ray background due to the absence of $\gamma$ rays 
originating from the excitation of the target nuclei. 
This high efficiency and low background conditions enabled us to observe 
the $2^+_1$ state in the very neutron-rich nucleus $^{74}$Ni in spite of 
the very low beam intensity of less than 1 count per second (cps). 

\section{Experiment}
\label{sec2}
The experiment was performed at the National Superconducting 
Cyclotron Laboratory (NSCL) at the Michigan State University 
using the large-acceptance A1900 fragment separator~\cite{Mor03} and 
the high-resolution S800 spectrograph~\cite{Baz03}.
A $^{86}$Kr primary beam accelerated to 140 MeV/nucleon 
by the Coupled Cyclotron Facility was directed onto a 399-mg/cm$^2$-thick 
$^9$Be fragmentation target at a typical intensity of 20 particle nA. 
The produced fragments were selected and separated by the A1900. 
The momentum acceptance was set to 2\%. 
At the second intermediate image of the A1900, an achromatic wedge degrader 
of 45-mg/cm$^{2}$-thick aluminum was placed to purify the secondary beam. 
Within a secondary cocktail beam, $^{74}$Ni was obtained with an average
intensity of 0.5 cps  while the total intensity of all transmitted fragments
amounted to 80 cps. 
 
The secondary beam was delivered to the analysis line of the S800 
spectrograph operated in focus mode; the beam was achromatically focused 
on the target. 
The particle identification was performed on an event-by-event basis  
by measuring the energy loss ($\Delta$$E$), 
magnetic rigidity ($B\rho$), and time-of-flight (TOF) of each nucleus.  
A Si-PIN detector with a thickness of 74~$\mu$m was used 
to measure $\Delta$$E$ of the incoming projectile beam at the entrance of the
S800 analysis line. The $B\rho$ value was obtained from the angle 
at the intermediate focal plane of the S800 analysis line 
measured by a pair of parallel-plate avalanche counters. 
The TOF was obtained from the time difference between the signals of plastic scintillators at the exit of the A1900 and 
the final focal plane of the S800 spectrograph. 

The secondary beam bombarded the liquid hydrogen target 
of thickness 210 mg/cm$^2$ to induce proton inelastic scattering.
The liquid hydrogen was contained in a cylindrical cell, a part of 
``CRYogenic ProTon and Alpha target system'' (CRYPTA)~\cite{Ryu05}. 
The target cell had aluminum entrance and exit windows 
with thicknesses of 0.22 mm and diameters of 30 mm. 
The contribution of the target windows to the inelastic scattering was estimated
to be 3\% of that by hydrogen. The average mid-target energy of $^{74}$Ni was 81
MeV/nucleon, which corresponds to center-of-mass energy of 80 MeV.  

\begin{figure}[t]
  \includegraphics[keepaspectratio, scale=0.8]{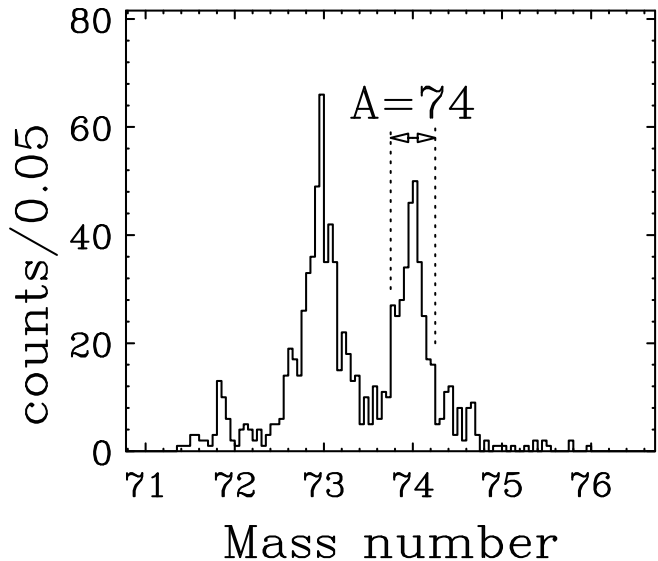}
 \caption{Mass distribution for Ni isotopes behind the secondary target 
 as measured with the S800 spectrograph. 
 The dotted lines indicate the region of the mass gate used in the present analysis.
 \label{fig:pi-ejectile}}
\end{figure}

The reaction residues were analyzed with the S800 spectrograph, 
whose magnetic field was set to transmit the 
(in)elastically scattered $^{74}$Ni ions. 
The particle identification was performed by combining 
$B\rho$, $\Delta$$E$, and TOF measured by the focal-plane 
detectors of the S800~\cite{JYur99}. 
The $B\rho$ value was obtained from the position of the ion at the final focal 
plane measured by two cathode readout drift chambers (CRDCs). 
An ionization chamber located downstream of the CRDCs provided the $\Delta$$E$ 
information. For the TOF, the flight time differences between the target
position and the final focal plane were used, the former being extracted from 
the time measured by the plastic scintillator at the exit of the A1900 
and $B\rho$ of the incident beam. 
The mass distribution for the Ni isotopes was obtained as shown 
in Fig.~\ref{fig:pi-ejectile} with a resolution of 0.6\% 
in full width at half maximum (FWHM). 
With this resolution, contaminations in the $^{74}$Ni mass gate 
set in the present analysis (indicated by the dotted lines in 
Fig.~\ref{fig:pi-ejectile}) are negligible. 

The $\gamma$ rays emitted from the excited nuclei were detected 
by a barrel array of NaI(Tl) scintillators with an inner diameter of 43
cm~\cite{Kal93,Per03} surrounding the liquid hydrogen target. 
The array consisted of 21 NaI(Tl) scintillator bars, 
each of which was 55.0 cm long and had a trapezoidal cross section  
with a height of 6.0 cm and upper and lower bases of 5.5 cm and 7.0 cm, 
respectively.  
The barrel covered a polar-angle range from 38 to 122 degrees with 
respect to the beam axis. This asymmetric geometry was adopted to enhance the
detection efficiency of $\gamma$ rays emitted in flight with forward-shifted
angular distributions due to the Lorentz boost. 
The signals were read out by two photomultiplier tubes optically coupled 
to both ends of each scintillator bar.
The energy of the $\gamma$ ray in the laboratory frame was obtained from the
geometrical average of the two signals from each pair of photomultiplier tubes. 
The detection position was obtained from the logarithm of the ratio of the two
signals. Using the position information, a Doppler-shift correction was applied
to the energy of the $\gamma$ rays emitted from the moving nucleus at a
velocity of $v/c\sim$0.4. Energy calibrations were performed with $^{22}$Na and
$^{137}$Cs standard sources. The intrinsic energy resolution was 
13\% (FWHM) for the 1275-keV $\gamma$ ray. The position was calibrated by using
a collimated $^{60}$Co source and a position resolution of 3 cm (FWHM) was
obtained.   

In order to extract the energy and intensity of the $\gamma$-ray transitions, 
the spectral shape expected for mono-energetic $\gamma$ rays from a moving 
source was computed with a Monte Carlo simulation 
using the {\scriptsize GEANT} code~\cite{GEANT}. 
The simulation took into account the geometries of the NaI(Tl) array, CRYPTA,
and the beam pipe, as well as the energy and position resolution and the energy
loss of the beam  in the target.
The overall energy resolution for the 1024 keV $\gamma$ rays emitted from 
$^{74}$Ni moving with $v/c \sim $0.4 was 18\% (FWHM). 
The photo-peak efficiency was calculated to be 16(2)\%, 
where the error is a systematic one estimated 
by comparing the measured $\gamma$-ray spectra of the standard sources 
with the simulated one. 

\section{Result}
\label{sec3}
\begin{figure}[t]
  \includegraphics[scale=0.8]{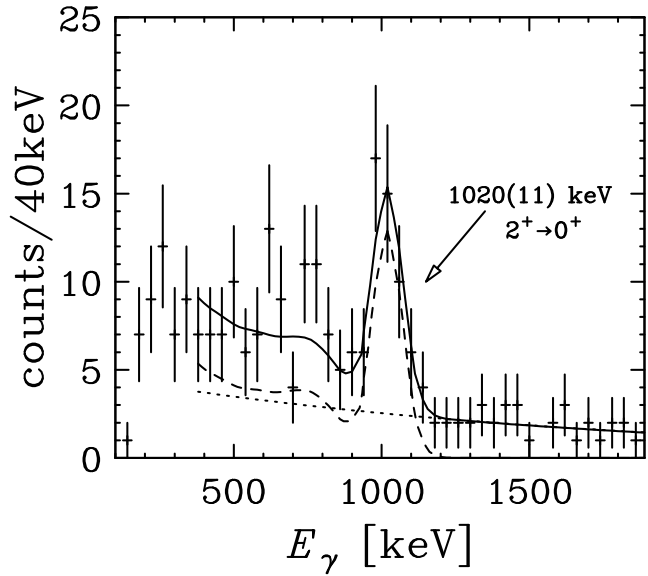}
 \caption{Doppler-shift corrected $\gamma$-ray spectrum for the proton 
inelastic scattering of $^{74}$Ni. The solid curve shows the best 
fit result, which consists of the simulated detector response for 
the 1020-keV $\gamma$ ray (dashed curve) and an exponential 
background (dotted curve). 
 \label{fig:74ni-gamma}}
\end{figure}

Figure~\ref{fig:74ni-gamma} shows the Doppler-shift corrected energy spectrum of
the prompt $\gamma$-rays detected in coincidence with the
inelastically-scattered $^{74}$Ni ions.  
A prominent peak is seen at 1020(11)~keV, 
which is ascribed to the $2_{1}^{+}\rightarrow 0_{\rm g.s.}^{+}$ transition 
from the selectivity of proton inelastic scattering. 
This energy is consistent with the value (1024(1)~keV) 
reported in the $\beta$-decay study of $^{74}$Co~\cite{Maz05}.
The peak intensity was extracted by fitting the spectrum. 
The best fit result is shown by the solid curve, 
which is composed of the simulated detector response for a 1020-keV $\gamma$ ray
(dashed curve) and an exponential background (dotted curve). 
In addition to the 1020-keV peak, a peak-like structure is visible  
at 786(30)~keV. The statistical significance of this peak was evaluated to be 
one standard deviation based on a fit that incorporated the 786-keV $\gamma$-ray
peak.  

The angle-integrated cross section for the $2_{1}^{+}$ excitation 
was determined from the yield of the $\gamma$ rays after 
considering possible feeding from higher excited states. 
In the observed spectrum,
the 786-keV peak is the only candidate for a feeding transition. Its
contribution amounts to 20(20)\% of the yield of the 1020-keV peak.  
Since the statistical significance is low and the coincidence 
relation with the $2_{1}^{+} \rightarrow 0_{\rm g.s.}^{+}$ transition 
is uncertain, the cross sections were derived with and without assuming 
the 786-keV peak as a feeding transition. Their average was adopted 
while the difference was taken into account as a systematic error. 

In addition, we considered the possible feeding from a 3$^{-}$ state, 
because proton inelastic scattering generally has a large cross section 
for the excitation to low-lying, collective 3$^{-}$ states. 
In the case of the stable Ni isotopes, 
the feeding transitions from the 3$^{-}$ state to the $2_{1}^{+}$ state have 
intensities of about 20$\sim$30\% 
of the direct excitation to the $2_{1}^{+}$ state~\cite{58-60Ni3-,62-64Ni3-}, 
and have energies around 2$\sim$3 MeV. 
Such a transition cannot be observed in the present measurement 
because of limited sensitivity. 
We therefore assumed the feeding contribution from 
the 3$^{-}$ state to be 25(5)\%, 
although the corresponding photo-peak was not observed. 
The $0_{\rm g.s.}^{+}\rightarrow 2_{1}^{+}$ excitation cross section 
was then determined to be 14(4) mb, 
where the quoted error includes both the statistical (20\%) 
and systematic uncertainties. 
The systematic error is attributed to the uncertainties 
in the efficiency of the $\gamma$-ray detection (10\%), 
the target thickness (3\%), 
and uncertainty of the feeding corrections (20\%). 

The deformation length $\delta^{p,p^{\prime}}$ for 
the $0_{\rm g.s.}^{+} \rightarrow 2_{1}^{+}$ transition 
was extracted from calculations based on  
the distorted wave theory using the {\scriptsize ECIS97} code~\cite{ECIS97}. 
We adopted two different approaches. 
The first one employed  
the global optical potential set KD02~\cite{Koning}.
By adopting the collective vibrational model, 
the transition potential was obtained from 
the derivative of the optical potential 
with an amplitude of $\delta^{p,p^{\prime}}$. 
The $\delta^{p,p^{\prime}}$ value was determined 
so that the calculated total inelastic scattering cross section reproduces 
the measured cross section. 
The $\delta^{p,p^{\prime}}$ value obtained in this way was 1.07(16)~fm, 
where the error includes the uncertainties of the experimental cross section.
The corresponding deformation parameter 
$\beta_{2}$ (=$\delta^{p,p^{\prime}} /R$ )
is 0.21(3) using a radius of $R=5.04$~fm 
which is taken from $R=r_0A^{1/3}$ with $r_{0}=1.2$~fm. 

In the second approach the optical potential and transition potential were
derived by folding the effective nucleon-nucleon interaction 
with the nucleon densities and transition densities, respectively, 
using the {\scriptsize MINC} code~\cite{MINC}. 
For the effective nucleon-nucleon interaction, 
the density-dependent effective interaction 
proposed by Jeukenne, Lejeune, and Mahaux (JLM)~\cite{JLM} was used 
with normalization factors of 
0.95 for the real and imaginary parts~\cite{JLMprm1,Take09}. 
The density distributions of protons and neutrons 
are of Woods-Saxon form with a common radius parameter of 4.51~fm, 
which was chosen to reproduce the root mean square (rms) radius 
of the matter distribution (4.03~fm) obtained from Hartree-Fock-Bogoliubov (HFB) 
calculations using the SkM$^{*}$ parameterization~\cite{QRPAden}.
The transition densities were obtained from the derivatives of the
proton and neutron densities. The $\delta^{p,p^{\prime}}$ value obtained with
this approach is 1.02(15)~fm.  The $\delta^{p,p^{\prime}}$ values extracted
using the global optical potential and folding model potential agree with each other. 
We take the average of 1.04(16)~fm as the adopted value for the subsequent
discussions. 

Due to the large neutron excess, $^{74}$Ni may have a neutron-skin structure 
where the neutron distribution has a larger radius than the proton 
distribution. Indeed, the above HFB calculation gives different rms radii for 
neutrons (4.13~fm) and protons (3.87~fm). In order to evaluate the effect of 
the possible neutron-skin structure, a microscopic calculation using 
Woods-Saxon distributions with different radius parameters for neutrons 
(4.66~fm) and protons (4.27~fm) was performed.  
These values were chosen to reproduce the rms radii obtained 
from the HFB calculation. The result is 1.03(16)~fm for $\delta^{p,p^{\prime}}$, 
which agrees with the one obtained without incorporating 
the neutron-skin effect. This shows that 
the possible neutron-skin structure does not have a significant 
effect on the present analysis. 

\section{Discussion}
\label{sec4}
\begin{figure}[t]
  \includegraphics[keepaspectratio, scale=0.7]{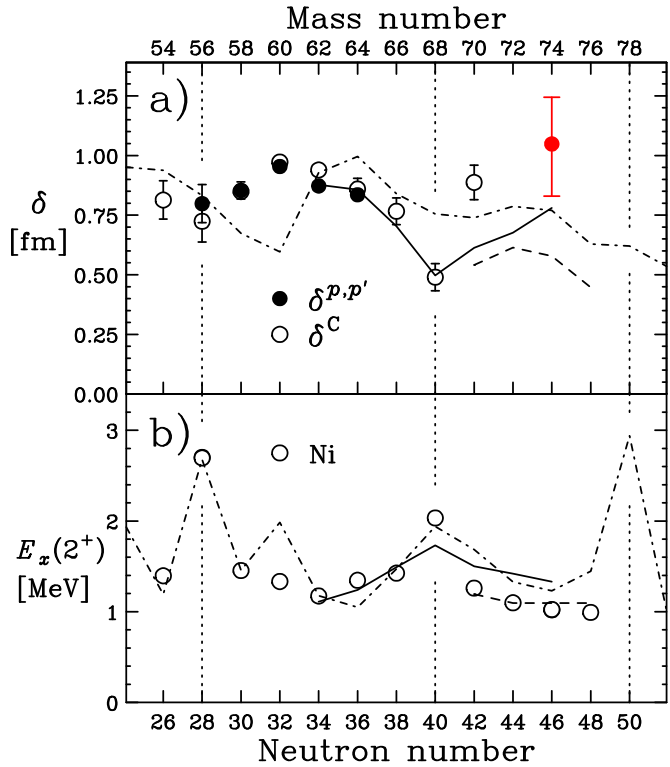} 
 \caption
{
(a) The $\delta$ values for Ni isotopes. 
   The open circles show $\delta^{\rm C}$ obtained from 
   $B({\rm E2})$ values~\cite{Yur04-1,Ram01,Sor02,Per06}.
   The filled circles show $\delta^{p,p^{\prime}}$ obtained from 
   the proton inelastic scattering~\cite{Kra94,Fab80}.
(b) The $E_{\rm x}(2^{+})$ systematics for Ni isotopes. 
In both figures the dashed and solid curves show the $\delta$ and $E_{\rm
  x}(2^{+})$ values from the shell-model calculations of Refs.~\cite{Lis04}
and~\cite{Sor02}, respectively. The dot-dashed curve shows the results of HFB+QRPA calculations~\cite{Tera06}.
\label{fig:z28-delta-theo} 
}
\end{figure}

The extracted $\delta^{p,p^{\prime}}$ value is shown in
Fig.~\ref{fig:z28-delta-theo}(a) together with those of stable Ni isotopes. 
The open circles indicate deformation lengths $\delta^{\rm C}$ 
obtained from $B({\rm E2})$ values using the following equation, 
\begin{eqnarray*}
\delta^{\rm C} &=& \beta_2^{\rm C}R \\
&=& \frac{4\pi}{3ZeR} \sqrt{B({\rm E2};0^+\rightarrow2^+)} 
\end{eqnarray*}
where $R=r_0A^{1/3}$ with $r_0=1.2$~fm. 
The $\delta^{p,p^{\prime}}$ value is sensitive to the quadrupole collectivity of 
both protons and neutrons, while $\delta^{C}$ is sensitive only to that of the
protons. The two quantities $\delta^{p,p ^\prime}$ and $\delta^{\rm C}$ are identical 
if the proton and neutron motions are the same. 
Indeed, the experimental values of $\delta^{p,p^\prime}$ and 
$\delta^{\rm C}$ are very close in the region where both values are available. 
In the following, we assume that the $\delta^{p,p^\prime}$ and $\delta^{\rm C}$ 
are close to each other also in the neutron-rich region, and 
both $\delta^{p,p^\prime}$ and $\delta^{\rm C}$ are referred 
to as $\delta$. 

The $\delta$ value for $^{74}$Ni obtained in the present study 
is about twice as large as that for $^{68}$Ni, and is comparable to or may even 
exceed that for $^{70}$Ni. This indicates that the enhancement of collectivity 
also persists in $^{74}$Ni, which is consistent with the trend of the $E_{\rm 
 x}(2^{+})$ values discussed in Section~1. 

The enhancement of quadrupole collectivity is inferred also for $^{72}$Ni, 
which could be estimated from the $B({\rm E2})$ value of $^{73}$Cu~\cite{Ste08}. 
It is suggested that the $3/2^-$ ground states and the low-lying 
$7/2^-$ states in the Cu isotopes in this region can be described 
in terms of a particle-core model. The evolution of the $B({\rm E2})$ values between
those states is  very similar to the behavior of the $B({\rm E2})$ values in the adjacent Ni
isotopes~\cite{Ste08}. The $B({\rm E2})$ value for $^{72}$Ni can then be estimated
from that of $^{73}$Cu, yielding a $\delta$ value of 1.09~fm, which is almost
the same as that of $^{74}$Ni. This is consistent with the above interpretation of
the enhanced collectivity in the Ni isotopes in this region. 

Figure ~\ref{fig:z28-delta-theo} compares the experimental $\delta$ and 
$E_{\rm x}(2^{+}_{1})$ values with theoretical ones. 
The dashed curve shows a shell-model calculation 
with the neutron configuration space of 
the $p_{3/2}, p_{1/2}, f_{5/2}$, and $g_{9/2}$ orbitals 
outside of a $^{56}$Ni inert core~\cite{Lis04}. 
In order to take into account the character of the $Z=28$ proton core,
which is rather soft~\cite{Hon02-c}, a large neutron effective 
charge of $e_{n}=1.0e$ was used. However, the calculation cannot reproduce the large 
$\delta$ values of $^{70}$Ni and $^{74}$Ni.
This means that the model space in this calculation is not large enough 
to account for the collectivity observed in the Ni isotopes. 
 
Results of a shell-model calculation with a larger model space reported 
in Ref.~\cite{Sor02} are shown by the solid curve. 
The model space comprises the  full $fp$-shell for protons and 
the $p_{3/2}, p_{1/2}, f_{5/2}$, and $g_{9/2}$ orbitals for neutrons. 
For effective charges, standard values of $1.5e$ and $0.5e$ 
were used for protons and for neutrons, respectively. 
The calculated $\delta$ values are larger than those of Ref.~\cite{Lis04} and 
the agreement with the experimental value has improved. 
However, the calculation still shows smaller $\delta$ values than the 
experiments for $^{70}$Ni and $^{74}$Ni. This underestimation of collectivity is 
consistent with the overestimation of the $E_{\rm x}(2^{+})$ values. 
This is in contrast to the results in the lighter region below neutron number $N=40$, 
where a good agreement is obtained for both $\delta$ and $E_{\rm x}(2^{+})$. 

Results of a calculation based on the HFB model and quasi-particle random phase
approximation (QRPA) using the SkM$^*$ effective interaction~\cite{Tera06} are
shown by the dot-dashed curve. This calculation also underestimates the $\delta$
values and  overestimates the $E_{\rm x}(2^{+})$ values in $^{70}$Ni and
$^{74}$Ni.  
The fact that these calculations fail to reproduce the large $\delta$ values 
for the Ni isotopes in the $N > 40$ region infers a modification of structure 
due to a mechanism characteristic for neutron-rich nuclei.  

One explanation for this phenomenon might lie in the softening of 
the $Z=28$ core. The shell gap at $Z=28$ is caused by the spin-orbit interaction 
and can easily be influenced by subtle changes of the potential 
due to unbalanced neutron and proton numbers. 
Indeed, the shell gap at $Z=28$ is predicted to be narrowed 
by the tensor interaction between protons 
and neutrons~\cite{Fra98,Ots05,Ots06,Ots10,Per06}. 
Since the tensor interaction is strongly attractive 
between $\pi f_{5/2}$ and $\nu g_{9/2}$ and strongly repulsive 
between $\pi f_{7/2}$ and $\nu g_{9/2}$, 
the energy gap between $\pi f_{5/2}$ and $\pi f_{7/2}$ is effectively 
reduced with increasing numbers of the valence neutrons in the $g_{9/2}$ orbital. 
As a result, the probability of proton excitations from the $f_{7/2}$ orbital into the 
upper orbitals increases, leading to the weakening of the shell closure.

The enhancement of collectivity could be due to neutrons, 
namely, to the enhanced probability of neutron excitations from the $g_{9/2}$ orbital 
into the $d_{5/2}$ orbital across the $N=50$ shell gap, which could occur if the
gap is narrowed.  
In the neighbouring isotopes, $N=50$ is known to be a good magic number
down to Zn isotopes~\cite{Pad05,Van07}. 
Therefore, this interpretation points to a weakening of the $N=50$ neutron
magicity characteristic in the neutron-rich Ni isotopes. 
In both scenarios, the collectivity could be more enhanced in the more neutron-rich
region and thus the doubly magic nature of $^{78}$Ni may be weakened. 
To further clarify the origin and mechanism of the enhanced collectivity 
observed in the present study, an investigation of nuclei closer to $^{78}$Ni 
is necessary. 

\section{Summary}
\label{sec5}
In summary, 
the neutron-rich nucleus $^{74}$Ni was studied 
by proton inelastic scattering in inverse kinematics 
at a center-of-mass energy of 80 MeV. 
The hitherto impossible measurement of this extremely exotic nucleus was 
realized by combining the high beam intensity available at NSCL with the
efficient use of a thick liquid hydrogen target (CRYPTA) in conjunction with
$\gamma$-ray spectroscopy to tag the inelastic excitation. The observation of
the $\gamma$-ray peak at 1020(11)~keV confirms the energy of the
$2^{+}\rightarrow0^{+}$ transition reported in an earlier $\beta$-decay study. 
The angle-integrated cross section for the excitation of the $2_1^+$ state 
was determined to be $\sigma=14(4)$~mb. From the comparison with the distorted 
wave theory, the $\delta$ value of 1.04(16)~fm was extracted.  
This large $\delta$ value suggests that the enhancement of
collectivity reported for $^{70}$Ni continues up to $^{74}$Ni.
The obtained $\delta$ value is larger than all values obtained by 
shell-model and HFB+QRPA calculations,  although both approaches reproduce
$\delta$ in the lighter Ni isotopes with $N < 40$.  
This infers a modification of the nuclear structure 
due to a mechanism characteristic for neutron-rich nuclei.  
The enhanced collectivity of $^{74}$Ni located
near the doubly-magic nucleus $^{78}$Ni 
may be an indication of the weakening of the magicity at $Z=28$ or $N=50$. 
In order to clarify the mechanism driving the enhancement in collectivity, 
a study of the structure of nuclei closer to $^{78}$Ni is needed. 

\section*{Acknowledgments}
We would like to thank the NSCL Cyclotron operators
for stable operation during the experiment.
The relocation of CRYPTA and the experiment are
supported by the U.S. National Science Foundation under 
Grants No. PHY-0110253, INT-0089581 
and by the Japan Society for the Promotion of Science.
We would like to thank Dr. J.~Terasaki for fruitful discussions.


\end{document}